\documentclass[aps,prd,preprintnumbers,superscriptaddress,nofootinbib,showpacs,twocolumn]{revtex4}%
\usepackage[dvips]{graphicx}
\usepackage{bm,latexsym,amsmath,amssymb,amsfonts,mathrsfs}
\usepackage{color}
\input{colordvi.tex}
\newcommand*{\D}{{\rm d}}
\newcommand*{\mpl}{M_{\rm Pl}}
\newcommand{\simgt}{\lower.5ex\hbox{$\; \buildrel > \over \sim \;$}}
\newcommand{\simlt}{\lower.5ex\hbox{$\; \buildrel < \over \sim \;$}}
\newcommand{\solM}{M_{\odot}}
\begin{document}

\title{Testing general scalar-tensor gravity and massive gravity with cluster lensing}


\author{Tatsuya~Narikawa}
\email[Email: ]{narikawa"at"resceu.s.u-tokyo.ac.jp}
\affiliation{Research Center for the Early Universe (RESCEU), Graduate School of Science,
The University of Tokyo, Tokyo 113-0033, Japan
}

\author{Tsutomu~Kobayashi}
\email[Email: ]{tsutomu"at"rikkyo.ac.jp}
\affiliation{Department of Physics, Rikkyo University, Toshima, Tokyo 175-8501, Japan
}

\author{Daisuke~Yamauchi}
\email[Email: ]{yamauchi"at"icrr.u-tokyo.ac.jp}
\affiliation{Institute for Cosmic Ray Research, The University of Tokyo, Kashiwa 277-8582, Japan
}

\author{Ryo~Saito}
\email[Email: ]{rsaito"at"yukawa.kyoto-u.ac.jp}
\affiliation{Yukawa Institute for Theoretical Physics, Kyoto University,
Kyoto 606-8502, Japan
}

\begin{abstract}
We explore the possibility of testing modified gravity exhibiting the Vainshtein mechanism
against observations of cluster lensing. We work in the most general scalar-tensor
theory with second-order field equations (Horndeski's theory), and derive static and spherically
symmetric solutions, for which the scalar field is screened below a certain radius.
It is found that the essential structure of the problem in the most general case
can be captured by the program of classifying Vainshtein solutions out of different
solutions to a quintic equation, as has been performed in the context of massive gravity.
The key effect on gravitational lensing is that the second derivative of the scalar field can
substantially be large at the transition from screened to unscreened regions, leaving
a dip in the convergence. This allows us to put observational constraints
on parameters characterizing the general scalar-tensor modification of gravity.
We demonstrate how this occurs in massive gravity as an example, and
discuss its observational signatures in cluster lensing.
\end{abstract}

\preprint{RESCEU-3/13, RUP-13-1, ICRR-Report 644-2012-33, YITP-13-8}
\maketitle

\section{Introduction}

The discovery of the accelerated expansion of the Universe~\cite{Riess:1998cb, Perlmutter:1998np}
has generated 
high interest in consistent long-distance modification of gravity. 
Although gravity might be modified so as to account for cosmic acceleration, 
modification must be suppressed at short distances
because general relativity is confirmed to very high accuracy in the solar system.
Thus, for a modified theory of gravity to be consistent, 
it is important to equip a mechanism to screen the effect of modification 
that could otherwise persist down to small scales.
A wide class of modified gravity theories can be
(at least effectively) described in terms of a scalar-tensor theory,
and the screening mechanisms can be understood
by inspecting the structure of the scalar-tensor Lagrangian.
Mainly two mechanisms have been proposed so far.
One is the chameleon mechanism~\cite{Khoury:2003aq},
in which a density-dependent effective potential is introduced
to make the scalar field sufficiently massive in a high-density environment.
This mechanism is utilized for example in $f(R)$ gravity~\cite{DeFelice:2010aj}.
Another mechanism is known as the Vainshtein effect\cite{Vainshtein:1972sx},
which operates in models with
nonlinear derivative interactions.
In such models, nonlinearity becomes large in the vicinity of a matter source
to suppress the scalar-mediated force.
This mechanism is incorporated for example into the DGP model~\cite{Dvali:2000hr, Lue:2005ya},
the Galileon model~\cite{Nicolis:2008in}, and massive gravity~\cite{Fierz:1939ix, Rubakov:2008nh}.

In this paper, we discuss how the Vainshtein effect works 
in Horndeski's scalar-tensor theory~\cite{Horndeski:1974wa, Deffayet:2011gz, Kobayashi:2011nu}.
Horndeski's theory is the most general theory
composed of a single scalar field and a metric
that gives rise to second-order field equations, and
therefore includes a large number of concrete models of modified gravity
as specific cases.
We present an algebraic master equation
that determines a gravitational field around
any static and spherically symmetric distribution of nonrelativistic matter.
The similar problem of deriving the Vainshtein solution in the cosmological
background
was addressed in Ref.~\cite{Kimura:2011dc}. There, however, complete analysis was hindered by
the complexity of the governing equations. In the case where the effect of
the cosmic expansion may be ignored, the relevant equations are simplified so that
the problem actually reduces to what has been done in the context of 
the Galileon model~\cite{Nicolis:2008in} and massive gravity~\cite{Wyman:2011mp, Chkareuli:2011te, Sjors:2011iv, Sbisa:2012zk}.
It is therefore possible to determine completely the conditions under which
the Vainshtein solutions are available.
Our solution thus obtained
is useful in testing modified gravity.

Among various cosmological observations, we are particularly interested
in cluster lensing, which can provide the convergence over a wide range of radii
by combining strong- and weak-lensing measurements 
(see Refs. \cite{Bartelmann:1999yn, Umetsu:2010us, Postman:2011hg}, and references therein).
The precise measurement of the structure of galaxy clusters 
offers us an important clue to test modified theories of
gravity~\cite{Lombriser:2011zw, Narikawa:2012tg}.
The first paper on gravitational lensing in the context of massive gravity is 
Ref. \cite{Wyman:2011mp}. The gravitational sector considered there is given by a subclass of 
Horndeski's theory. However, Ref. \cite{Wyman:2011mp} considered nonminimal and disformal coupling 
to matter, while we assume that matter is minimally coupled to gravity. 
Working in Horndeski's theory,
one can examine modified gravity endowed with 
the Vainshtein screening mechanism in a generic manner.
In this paper, we compute the lensing convergence of galaxy clusters
in Horndeski's theory based on the general spherically symmetric solution.
As a specific example, we demonstrate how nonlinear massive gravity proposed recently
in Ref.~\cite{deRham:2010kj} is included in our analysis, and
explore the possibility of detecting the signatures of modified gravity
in the context of massive gravity.

This paper is organized as follows:
In the next section, we define Horndeski's Lagrangian
and explain our procedure to obtain static and spherically symmetric solutions.
Then, in Sec.~\ref{sec:eq},
we derive the Vainshtein solution and present the conditions
for the existence of the solution.
We show the relation between Horndeski's theory and
the decoupling limit of massive gravity in Sec.~\ref{sec:proxy}.
In Sec.~\ref{sec:lens} we study the effect of modification of gravity
on gravitational lensing and explore its implications for cluster lensing observations.
Section~\ref{sec:concl} is devoted to conclusions.

\section{Static and spherically symmetric configurations in Horndeski's theory}
\label{sec:config}

We consider the most general scalar-tensor theory
constructed by Horndeski.
The action we study is given by
\begin{eqnarray}
S=S_{\rm GG}\left[g_{\mu\nu}, \phi\right]
+S_{\rm m}\left[g_{\mu\nu},\psi_{\rm m}\right],
\end{eqnarray}
where $S_{\rm m}$ is the action for matter $\psi_{\rm m}$, which is 
assumed to be minimally coupled to gravity.
The first term is composed of the four Lagrangians,
\begin{eqnarray}
S_{\rm GG}=\int\D^4 x\sqrt{-g}\left({\cal L}_2
+{\cal L}_3+{\cal L}_4+{\cal L}_5\right),
\end{eqnarray}
where
\begin{eqnarray}
{\cal L}_2&=&K(\phi, X),\\
{\cal L}_3&=&-G_3(\phi, X)\Box\phi,\\
{\cal L}_4&=&G_4(\phi, X)R
+G_{4X}\left[(\Box\phi)^2-(\nabla_\mu\nabla_\nu\phi)^2\right],
\\
{\cal L}_5&=&
G_5(\phi, X)G^{\mu\nu}\nabla_\mu\nabla_\nu\phi
-\frac{1}{6}G_{5X}\bigl[(\Box\phi)^3
\nonumber\\&&\quad
-3\Box\phi(\nabla_\mu\nabla_\nu\phi)^2
+2(\nabla_\mu\nabla_\nu\phi)^3\bigr],
\end{eqnarray}
with $X:=-g^{\mu\nu}\partial_\mu\phi\partial_\nu\phi/2$.
The second term is the action for matter fields.
Here, $R$ is the Ricci scalar, $G_{\mu \nu}$ is the Einstein tensor, 
and $K$ and $G_i$ are arbitrary functions of $\phi$ and $X$.
Here we have written the Lagrangians in
the equivalent form called the generalized Galileon rather than
the original one.
In this paper, we use notation such as $G_{4X}:=\partial G_4/\partial X$
and $G_{3\phi}:=\partial G_3/\partial \phi$.

We are interested in spherical overdensities on subhorizon scales,
so that we will neglect the effect of cosmic expansion.
In Ref.~\cite{Kimura:2011dc},
the Vainshtein mechanism in Horndeski's theory has been investigated
taking into account the background evolution of the scalar field, $\phi=\phi_0(t)$,
and it was shown that
Newton's ``constant'' evolves in time through its dependence on
$\phi_0(t)$ and $X_0(t):=\dot\phi_0^2/2$
(see also Ref.~\cite{Babichev:2011iz} for a similar argument).
However, the time variation of Newton's ``constant''
is strongly constrained from experiments~\cite{Williams:2004qba}.
This leads us to assume that the scalar field has a negligible time dependence
at least during the relevant period, $\phi_0\simeq$ const and $X_0\simeq 0$.

Our background solution is thus taken to be
\begin{eqnarray}
\D s^2=\eta_{\mu\nu}\D x^\mu\D x^\nu,
\quad
\phi=\phi_0={\rm const},\quad X=0.\label{BG}
\end{eqnarray}
In order for the theory to admit this solution,
we require that
$K(\phi_0, 0)=0$ and $K_\phi(\phi_0, 0)=0$.

Spherically symmetric perturbations
produced by a nonrelativistic matter lump
on top of the
background~(\ref{BG}) can be written as
\begin{eqnarray}
\D s^2&=&
-[1+2\Phi(r)]\D t^2+[1-2\Psi(r)]\delta_{ij}\D x^i\D x^j,
\\
\phi &=& \phi_0+\varphi(r),
\end{eqnarray}
where $r$ is the usual radial coordinate, $r^2=x^2+y^2+z^2$.
The time-time component of the gravitational field equations is
\begin{eqnarray}
&&G_4\frac{(r^2\Psi')'}{r^2}-G_{4\phi}\frac{(r^2\varphi')'}{2r^2}
-(G_{4X}-G_{5\phi })\frac{[r(\varphi')^2]'}{2r^2}
\nonumber\\&&\quad
+G_{5X}\frac{[(\varphi')^3]'}{6r^2}
=- \frac{1}{4}T_{t}^{\;t},\label{F1}
\end{eqnarray}
while the space-space component reduces to
\begin{eqnarray}
2G_4\left(\Psi'-\Phi'\right)
-2G_{4\phi}\varphi'
-(G_{4X}-G_{5\phi})\frac{(\varphi')^2}{r}=0,\label{F2}
\end{eqnarray}
where a prime stands for differentiation with respect to $r$.
Finally, from the scalar-field equation of motion we obtain
\begin{eqnarray}
&&(K_X-2G_{3\phi})\frac{(r^2\varphi')'}{r^2}-2(G_{3X}-3G_{4\phi X})\frac{[r(\varphi')^2]'}{r^2}
\nonumber\\&&
+2G_{4\phi}
\frac{[r^2(2\Psi -\Phi)']'}{r^2}
+4(G_{4X}-G_{5\phi})\frac{[r\varphi'(\Psi'-\Phi')]'}{r^2}
\nonumber\\&&
+2\left(G_{4XX}-\frac{2}{3}G_{5\phi X}\right)
\frac{[(\varphi')^3]'}{r^2}+2G_{5X}\frac{[(\varphi')^2\Phi']'}{r^2}
\nonumber\\&&
=-K_{\phi\phi}\varphi.\label{F3}
\end{eqnarray}
Here, all the functions in the coefficients
are evaluated at $\phi=\phi_0$ and $X=0$.
From now on, we will ignore the mass term $K_{\phi\phi}$,
because we focus only on modified gravity endowed with the Vainshtein mechanism.

One sees that Eqs.~(\ref{F1}) and~(\ref{F3})
can be integrated once to give algebraic equations for $\varphi',\;\Phi',\;\Psi'$.
In doing so it is convenient to use the enclosed mass defined as
\begin{eqnarray}
M(r):=4\pi\int^r_0\left(-T_t^{\;t}\right)r^2\D r.
\end{eqnarray}
The resultant equations coincide with those derived
from taking the limit $\phi=\phi_0=$ const
and $X=0$
in the result of Ref.~\cite{Kimura:2011dc}.

Let us introduce six dimensionless parameters,
$\xi, \eta, \mu, \nu, \alpha$, and $\beta$,
as well as the Planck mass $\mpl$ and
a new mass scale $\Lambda$, to rearrange and simplify the expressions.
Those dimensionless quantities are
related to the coefficients in the above equations as
\begin{eqnarray}
G_4=\frac{\mpl^2}{2},
\quad
G_{4\phi}=\mpl \xi,
\end{eqnarray}
and
\begin{eqnarray}
K_X-2G_{3\phi}&=&\eta,
\\
-G_{3X}+3G_{4\phi X}&=&\frac{\mu}{\Lambda^3},
\\
G_{4X}-G_{5\phi}&=&\frac{\mpl}{\Lambda^3}\alpha,
\\
G_{4XX}-\frac{2}{3}G_{5\phi X}&=&\frac{\nu}{\Lambda^6},
\\
G_{5X}&=&-\frac{3\mpl}{\Lambda^6}\beta.
\end{eqnarray}
We also define
\begin{eqnarray}
x(r)=\frac{1}{\Lambda^3}\frac{\varphi'}{r},
\quad
A(r)=\frac{1}{\mpl\Lambda^3}\frac{M(r)}{8\pi r^3},
\end{eqnarray}
both of which are dimensionless.

Now the gravitational field equations~(\ref{F1})
and~(\ref{F2}) reduce to
\begin{eqnarray}
\frac{\mpl}{\Lambda^3}\frac{\Phi'}{r}
&=&-\xi x+\beta x^3+A(r),\label{Phi-eq}
\\
\frac{\mpl}{\Lambda^3}\frac{\Psi'}{r}
&=&\xi x+\alpha x^2+\beta x^3+A(r).\label{Psi-eq}
\end{eqnarray}
Substituting Eqs.~(\ref{Psi-eq}) and (\ref{Phi-eq}) to
the scalar-field equation of motion~(\ref{F3}), we arrive at
\begin{widetext}
\begin{eqnarray}
P(x, A):=\xi A(r)+\left(\frac{\eta}{2}+3\xi^2\right)x
+\left[\mu+6\alpha\xi-3\beta A(r)\right]x^2
+\left(\nu+2\alpha^2+4\beta\xi\right)x^3-3\beta^2x^5=0.\label{basic-equation}
\end{eqnarray}
\end{widetext}
Solving the algebraic equation $P(x, A)=0$ for $x$,
one obtains the radial profile of $x$
in terms of $A=A(r)$. It is then straightforward to
determine the two metric potentials by using Eqs.~(\ref{Phi-eq})
and~(\ref{Psi-eq}).
Note in passing that if $\xi=0$
then we have a trivial solution $x(r)=0$ which is not interesting.
In the rest of the paper we therefore assume that $\xi\neq 0$.

Since there is a sufficient number of parameters,
at this stage the coefficients of the polynomial $P(x, A)$
are free in principle.
However, it is important to note that
the structure of $P(x, A)$
in the most general case
is still essentially the same as the corresponding equation
in massive gravity:
it is quintic and the matter source term $A$
appears only in the zeroth-order and quadratic terms.
This structure allows us to
proceed following closely
the previous analysis in massive gravity~\cite{Sjors:2011iv, Sbisa:2012zk}.

\section{The quintic and cubic equations}
\label{sec:eq}

In this section, we solve the equation $P(x,A)=0$
for $A\gg 1$ and $A\ll 1$,
and single out a solution appropriate for our current purpose in each domain.
We then derive the conditions under which
the two solutions are matched smoothly in an intermediate region.
The procedure here is basically the same as
that of Ref.~\cite{Sbisa:2012zk}.
The region far from the source corresponds to $A\ll 1$,
while it is assumed that $A\gg 1$ in the vicinity of the source,
where the Vainshtein mechanism is expected to operate.
It is therefore appropriate to define the Vainshtein radius $r_{\rm V}$
by
\begin{eqnarray}
A(r_{\rm V})=1.\label{def-Vain}
\end{eqnarray}

In the outer region ($A\ll 1$), there is always a decaying solution,
\begin{eqnarray}
x\approx x_{\rm f}:= -\frac{2\xi A(r)}{\eta+6\xi^2},\label{outsol}
\end{eqnarray}
which is obtained by neglecting the nonlinear terms in $P(x, A)$.
We are interested only in this solution, because
the other solutions, if they exist, do not correspond to
an asymptotically flat spacetime.

The stability of the solution in the linear regime
can be studied by using the truncated action
\begin{eqnarray}
S=\int \D^4x\sqrt{-g}
\left[\frac{\mpl^2}{2}\left(1+2\xi\frac{\phi-\phi_0 }{\mpl}\right)R+\eta X\right].
\end{eqnarray}
Working in this action, one can derive
the stability condition easily in the same way as in the
Brans-Dicke theory:
{\em in the Einstein frame},
the kinetic term for small fluctuations
has the right sign provided that
\begin{eqnarray}
\eta+6\xi^2>0.\label{stability-linear}
\end{eqnarray}
We require the condition~(\ref{stability-linear})
for the stability of the solution $x\approx x_{\rm f}$.
The same condition can also be derived from $\partial P(x_{\rm f}, A)/\partial x>0$.
(See Appendix~A for further details.)

Let us turn to identifying the desired inner solution.
The inner solution is different depending on whether
$\beta\neq 0$ or $\beta=0$, because the structure of $P(x, A)$
for $A\gg 1$ is crucially different.

\subsection{$\beta\neq 0$}

In the inner region ($A\gg 1$), Eq.~(\ref{basic-equation})
reduces to
\begin{eqnarray}
P(x, A)\approx \xi A-3\beta Ax^2-3\beta^2x^5 \approx 0.\label{large-A}
\end{eqnarray}
The behavior of the solution to this equation
depends on the sign of $\xi\beta$.\\

(i) $\xi\beta<0$.
%
In this case, the second and the third terms in Eq.~(\ref{large-A})
balance, so that the solution is
\begin{eqnarray}
x^3\approx -\frac{A}{\beta}.
\end{eqnarray}
For this inner solution we find
$\Psi'/r\propto A^{2/3}$ and $\Phi'/r\propto A^{1/3}$,
which is not at all standard gravity.
We therefore discard this possibility.

(ii) $\xi\beta>0$.
%
In this case, we have the same solution as above,
$x^3\approx -A/\beta$,
which is to be discarded for the same reason.
We have another solution,
\begin{eqnarray}
x\approx x_{\pm}
:= \pm\sqrt{\frac{\xi}{3\beta}}={\rm const},\label{insol}
\end{eqnarray}
for which the first and the second terms
in Eq.~(\ref{large-A}) balance.
For this latter solution we have the correct Newtonian behavior:
$\Psi'/r\simeq \Phi'/r\propto A$.
This is therefore the Vainshtein solution we are looking for.\\


Having thus identified the desired solutions in the
inner and outer regions, we consider the matching of the two.
For simplicity we focus on the case of $\xi>0$.
The solution matching in the case of $\xi<0$ can also be done
essentially in the same way as follows.

For $\xi>0$ (and hence $\beta>0$),
the outer solution $x=x_{\rm f}<0$ can be matched only to
the inner solution $x=x_{-}<0$ smoothly.\footnote{Since
$\partial P(x_\pm, A)/\partial x=-6\beta x_{\pm}$,
the solution $x=x_-$ is stable, but the other one, $x=x_+$, is not.}
The smooth matching of the two solution is possible
if and only if {\em $P(x, A)=0$ has a single root in $(x_-, 0)$ for any $A>0$.}
Since $P(0, A)=\xi A>0$, if 
\begin{eqnarray}
P(x_-, A)<0,\label{cond1}
\end{eqnarray}
the intermediate value theorem
guarantees the existence of at least one root in $(x_-, 0)$.
Note that $P(x_-, A)$ does not in fact depend on $A$.
Using the intermediate value theorem again in $(-\infty, x_{-})$ and $(0,\infty)$, 
it can be shown that one or three roots exist in general in $(x_{-},0)$.

Let $x_*$ and $A_*$ be the solution to
\begin{eqnarray}
\frac{\partial P(x_*, A_*)}{\partial x}=0,
\quad
\frac{\partial^2 P(x_*, A_*)}{\partial x^2}=0.\label{exist2}
\end{eqnarray}
If such $x_*\in (x_-, 0)$ and $A_*>0$ do not exist, it is obvious that
there is only a single root in $(x_-, 0)$.
In this case, the smooth matching is possible.
If such $x_*$ and $A_*$ exist, $P(x, A)=0$ would
in general have three roots in $(x_-, 0)$
for some interval of $A$.
Since $P(x, A)=0$ for $A\gg 1$
[Eq.~(\ref{large-A})] has a single root in $(x_-, 0)$,
two of the three roots in $(x_-,0)$ disappear as $A$ increases.
For parameters satisfying $P(x_*, A_*)>0$, the solution
corresponding to $x_{\rm f}$ disappears as $A$ increases,
implying that the smooth matching is impossible in this case.
However, if
\begin{eqnarray}
P(x_*, A_*)< 0,\label{cond2}
\end{eqnarray}
the solution corresponding to $x_{\rm f}$
remains and hence the smooth matching is still possible.
Indeed, there is a single root for any $A>0$ in this case
because a simple manipulation shows
that $P(x_{\rm e}(A), A)$ increases with increasing $A$,
where $x_{\rm e}(A)$ is the locus of extrema in $(x_-, 0)$.

Summarizing, the outer and inner solutions can be matched smoothly
provided that Eq.~(\ref{cond1}) and either of the following two
conditions are satisfied:
(i) $x_*\in (x_-, 0)$ and $A_*>0$ do not exist satisfying Eq.~(\ref{exist2});
(ii) such $x_*\in (x_-, 0)$ and $A_*>0$ exist, but they satisfy Eq.~(\ref{cond2}).

\subsection{$\beta=0$}

Here again we focus on the case of $\xi>0$ for simplicity.
The problem reduces to solving the cubic equation,
$P(x, A)=0$, where now
\begin{eqnarray}
P(x, A)&\to&\xi A+\left(\frac{\eta}{2}+3\xi^2\right)x
+\left(\mu+6\alpha\xi\right)x^2
\nonumber\\&&
+\left(\nu+2\alpha^2\right)x^3.
\end{eqnarray}
The inner solution is
\begin{eqnarray}
x^3\approx x_{\rm i}^3:= -\frac{\xi A}{\nu+2\alpha^2}.
\end{eqnarray}
(We assume that $\nu +2\alpha^2 \neq 0$.)
This solution shows the correct Newtonian behavior:
$\Psi'/r\simeq \Phi'/r\propto A$.

The inner and outer solutions can be matched smoothly
only for $\nu+2\alpha^2>0$, which is also required from stability
of the inner solution.
Smooth matching also requires that
there is a single root in $x<0$ for any $A>0$, or, equivalently,
that
$P(x, A)$ has no local extrema in $x<0$.
The two local extrema are in $x>0$ provided that
\begin{eqnarray}
\mu+6\alpha \xi<0, \label{c3-1}
\end{eqnarray}
where we used the condition~(\ref{stability-linear}).
Otherwise, one must require that
the discriminant of $\partial P(x, A)/\partial x$ is negative, {\em i.e.,}
\begin{eqnarray}
\left(\nu+2\alpha^2\right)\left(\eta+6\xi^2\right)
\ge \frac{2}{3}\left(\mu+6\alpha\xi\right)^2 ,
\;\; \mu+6\alpha\xi\ge 0.\label{c3-2}
\end{eqnarray}
Summarizing, smooth matching is possible if
Eq.~(\ref{c3-1}) or Eq.~(\ref{c3-2}) is satisfied.

\section{Decoupling limit of massive gravity}\label{sec:proxy}

Let us confirm that the conditions for smooth matching
indeed reproduce the previous result obtained in the
context of massive gravity~\cite{Sjors:2011iv, Sbisa:2012zk}.
To do so, we start with finding out the concrete form
of $K, G_3, G_4, G_5$ corresponding to the decoupling limit of massive gravity.
The correspondence can be seen more clearly
if we move to the covariantized version of the decoupling limit Lagrangian,
{\em i.e.,} the ``proxy theory'' proposed in Ref.~\cite{deRham:2011by}.

It turns out that the proxy theory
corresponds to
\begin{eqnarray}
&&K=0=G_3,
\quad
G_4=\frac{\mpl^2}{2}+\mpl \phi +\frac{\mpl}{\Lambda^3}\alpha X,
\nonumber\\&&
G_5=-3\frac{\mpl}{\Lambda^6}\beta X.\label{proxy-G}
\end{eqnarray}
In massive gravity, the strong coupling scale $\Lambda$ is given by $\Lambda=(m^2 \mpl)^{1/3}$,
where $m$ is the graviton mass.

Since the proxy theory contains the Riemann dual tensor while
the Lagrangian of the generalized Galileon not, one may wonder
how the former is included in the latter.
Actually, $G_5\propto X$ corresponds to the term containing the Riemann dual tensor
in the proxy theory. The easiest way to verify this is
to compare the field equations of the two theories.

From Eq.~(\ref{proxy-G}) one finds
\begin{eqnarray}\label{eq:PSpace}
\eta=\mu=\nu=0,
\quad
\xi=1,
\quad
\alpha\neq 0, \quad \beta\neq 0,
\end{eqnarray}
so that the parameter space collapses to a two-dimensional space.
The inner solution $x_-$ exists only for $\beta>0$ and is given by
$x_-=-1/\sqrt{3\beta}$.
Let us define $\zeta:=\sqrt{\beta}/\alpha$.
Then, the condition~(\ref{cond1}) reads
\begin{eqnarray}
P(x_-, A)=\frac{2}{3}\frac{x_-}{\zeta^2}
\left(1-3\sqrt{3}\zeta+6\zeta^2\right)<0.
\end{eqnarray}
Solving the equation $\partial_x P(x_*, A_*)-x_*\partial_x^2 P(x_*, A_*)=0$,
which does not in fact depend on $A_*$, one finds
\begin{eqnarray}
x_*=\frac{1}{\sqrt{5}}\frac{x_-}{|\zeta|}
\left[1+2\zeta^2
-\left(1+4\zeta^2-11\zeta^4\right)^{1/2}\right]^{1/2}.
\end{eqnarray}
This exists if
\begin{eqnarray}
|\zeta|\le \sqrt{\frac{2+\sqrt{15}}{11}}\simeq 0.73.
\end{eqnarray}
The equation $P(x, A)=0$ has three roots in $(x_-, 0)$ for some interval of $A$ if
\begin{eqnarray}
P(x_*, A_*)>0
\quad\Leftrightarrow\quad 0<\zeta<\sqrt{\frac{5+\sqrt{13}}{24}}\simeq 0.6.
\end{eqnarray}
Therefore, smooth matching
of the asymptotically flat solution and the Vainshtein solution
is possible provided that
\begin{eqnarray}
\alpha <0\quad{\rm or}\quad
\frac{\sqrt{\beta}}{\alpha} \ge \sqrt{\frac{5+\sqrt{13}}{24}}.
\end{eqnarray}
Thus, we have confirmed that
the previous result~\cite{Sjors:2011iv, Sbisa:2012zk} is reproduced.\footnote{Note that
our notation is different from those in~\cite{Sjors:2011iv, Sbisa:2012zk}.
In particular, $\alpha_{\rm ours} = -\alpha_{{\rm Sbisa}\, et\, al.}$.}

\begin{figure}[tbp]
  \begin{center}
    \includegraphics[keepaspectratio=true,height=80mm]{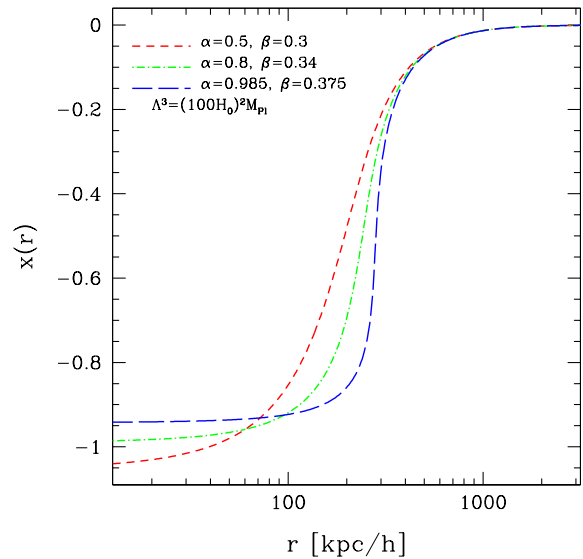}
  \end{center}
  \caption{
The profile of $x$
  as a function of the radial coordinate $r$.
  The curves are plotted for $(\alpha, \beta)=(0.5, 0.3)$ (dotted red line), $(0.8, 0.34)$ (dot-dashed green line), 
  and $(0.985, 0.375)$ (dashed blue line), respectively. 
As a halo density profile
we adopt the NFW model with 
  $M_{\rm vir}=1.34\times10^{15} \solM/h$ and $c_{\rm vir}=13.8$.
}%
\label{fig:xr}
\end{figure}

\begin{figure}[tbp]
  \begin{center}
    \includegraphics[keepaspectratio=true,height=80mm]{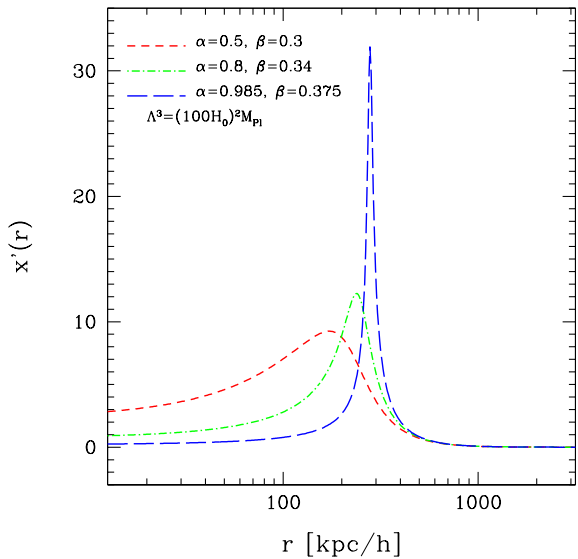}
  \end{center}
  \caption{
The radial derivative of $x$ as a function of $r$.
  Parameters and definitions of curves
  are the same as in
  Fig.~\ref{fig:xr}.
}%
\label{fig:dxdr}
\end{figure}

\section{Gravitational lensing in modified gravity}\label{sec:lens}

In this section, we are going to relate our spherically symmetric
solution to gravitational lensing observations.
To do so, it is instructive to begin
with seeing the typical behavior of the Vainshtein solution in massive gravity,
adopting the Navarro-Frenk-White (NFW)
halo density profile~\cite{Navarro:1995iw, Navarro:1996gj}
for the source $\rho(r):=-T_t^{\;t}$.
(See Appendix~B for the detailed description of halo density profiles.)
Figures~\ref{fig:xr} and~\ref{fig:dxdr} show
the profile of $x$ and its derivative, respectively,
as a function of the radial coordinate $r$
for different values of $\alpha$ and $\beta$.
The fiducial parameters of the NFW model we use are
$M_{\rm vir}=1.34\times 10^{15}\,\solM/h$ and $c_{\rm vir}=13.8$\,,
which correspond to $\rho_{\rm s}=7.16\times 10^4\,\rho_{\rm cr,0}$ and
$r_{\rm s}=145\,{\rm kpc}/h$, respectively.
The strong coupling scale is taken to be $\Lambda^3 = (100H_0)^2 \mpl = (46.4\,{\rm km})^{-3}$.
Then, the Vainshtein radius determined from Eq.~\eqref{def-Vain} is $r_{\rm V}=209\,{\rm kpc}/h$.
(As the parameters characterizing the profile
we choose to use the virial cluster mass $M_{\rm vir}$ and
the concentration parameter $c_{\rm vir}$ rather than $\rho_{\rm s}$ and $r_{\rm s}$.)
One can see that $x(r)$ can have a sharp transition
from outer to inner solutions, depending on the parameters of the theory,
which leads to a peak in $x'(r)$.
This occurs at around the Vainshtein radius.

Having seen the typical behavior of the radial profile $x(r)$,
we now move to investigate how the lensing signal is modified in massive gravity.
We assume that the background evolution of the Universe does not deviate much from
conventional cosmology and use the $\Lambda$CDM background with
$\Omega_{\rm m}=0.3$\,, $\Omega_\Lambda =0.7$\,, and $h=0.7$. 
The background metric~(\ref{BG}) is understood to define
the physical coordinates at the location of the lens object.

The basic quantity in gravitational lensing is the convergence, $\kappa$,
which is expressed
in terms of the sum of the two metric potentials $\Phi_+:=(\Phi+\Psi)/2$ as
\begin{eqnarray}
\kappa
= \int_0^{\chi_{\rm S}} \D\chi
\frac{(\chi_{\rm S}-\chi )\chi}{\chi_{\rm S}} \Delta_\perp \Phi_+,
\end{eqnarray}
with
$\chi$, $\chi_{\rm S}$, and $\Delta_\perp$ being
the comoving angular diameter distance,
the comoving distance between the observer and the source,
and the comoving transverse Laplacian, respectively.
Using the thin lens approximation, 
we can rewrite the convergence as
\begin{eqnarray}
\kappa \simeq
\frac{(\chi_{\rm S}-\chi_{\rm L})\chi_{\rm L}}{\chi_{\rm S}}
\int_0^{\chi_{\rm S}} \D\chi\, \Delta
\Phi_+, \label{kappa}
\end{eqnarray} 
where $\chi_{\rm L}$ is the comoving distance between the observer and the lens object
and $\Delta$ is the comoving three dimensional Laplacian.
Let us now introduce a new spatial coordinate as
$Z=a_{\rm L}(\chi-\chi_{\rm L})$, 
whose origin is located at the center of the lens object.
The projected radius is written as $r_\perp=a_{\rm L}\chi_{\rm L}\theta$,
where $\theta$ is the polar angle from the axis connecting the observer and the lens object,  
and $a_{\rm L}$ is the scale factor at the lens object.
In terms of these,
the convergence (\ref{kappa}) can be written as
\begin{eqnarray}
\kappa(\theta)
&=&\frac{2(\chi_{\rm S}-\chi_{\rm L})\chi_{\rm L}a_{\rm L}}{\chi_{\rm S}}
\int_0^\infty \D Z \frac{\Delta}{a_{\rm L}^2} \Phi_+(r),
\label{defkappa}
\end{eqnarray}
where $r=\sqrt{r_\perp^2+Z^2}$.
Using Eqs.~(\ref{Phi-eq})
and~(\ref{Psi-eq}),
we find that in Horndeski's theory
\begin{eqnarray}
\frac{\Delta}{a_{\rm L}^2} \Phi_+ (r) 
&=& \frac{1}{r^2}\frac{\D}{\D r}\left[r^2\Phi_+'(r)\right]
\nonumber\\
&=&\frac{\Lambda^3}{\mpl}
\frac{[\left(\alpha x^2+2\beta x^3+2A\right)r^3]'}{2r^2}.
\end{eqnarray}

\begin{figure}[tbp]
  \begin{center}
    \includegraphics[keepaspectratio=true,height=80mm]{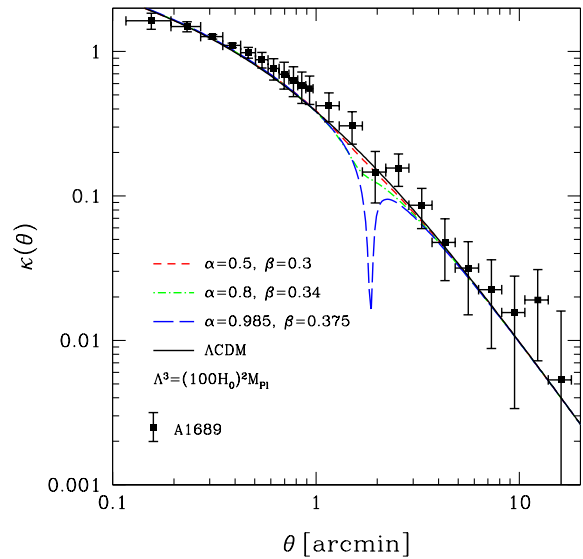}
  \end{center}
  \caption{The lensing convergence $\kappa$
  as a function of $\theta$ 
  for different values of the parameters of the theory.
  In these plots,
  the NFW profile is used with $M_{\rm vir}=1.34\times 10^{15}\,\solM/h$ and $c_{\rm vir}=13.8$.
  Parameters and definitions of the curves are the same as in Fig.~\ref{fig:xr}.
  The points with the error bars represent the observational data 
  for the high-mass cluster A1689
  provided by Umetsu {\it et al.}~\cite{Broadhurst:2004hu, Umetsu:2007pq, Umetsu:2010rv, Umetsu:2011ip, Postman:2011hg}.
  }
  \label{fig:kappa}
\end{figure}

\begin{figure}[tbp]
  \begin{center}
    \includegraphics[keepaspectratio=true,height=80mm]{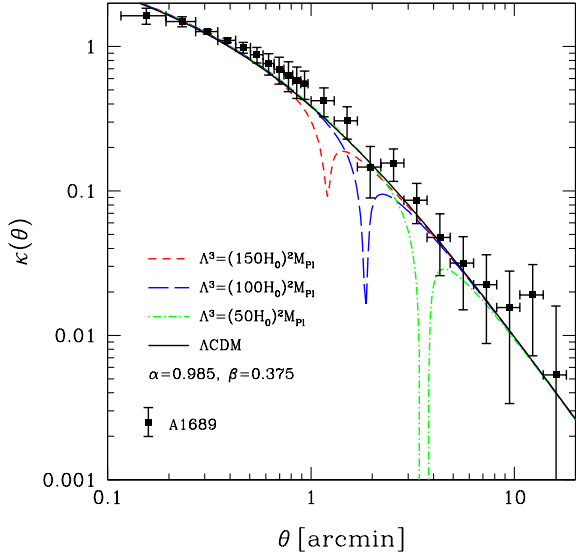}
  \end{center}
  \caption{
  The convergence $\kappa$ as a function of $\theta$
  for different strong coupling scales.
 The curves correspond to $\Lambda^3=(150H_0)^2\mpl$ (dotted red), $\Lambda^3=(100H_0)^2\mpl$ (dashed blue),
 $\Lambda^3=(50H_0)^2\mpl$ (dot-dashed green), and $\Lambda$CDM (black solid), respectively.
 We take $\alpha=0.985$ and $\beta=0.375$.
}%
  \label{fig:kappa.Lamb_ch}
\end{figure}

Figure~\ref{fig:kappa} shows the lensing convergence for the NFW profile
with different choices of the parameters of the theory.
The points with the error bars indicate the observational data 
for the high-mass cluster A1689 provided
by Umetsu {\it et al.}~\cite{Broadhurst:2004hu, Umetsu:2007pq, Umetsu:2010rv, Umetsu:2011ip, Postman:2011hg}.
An interesting feature observed in Fig.~\ref{fig:kappa} is that
a dip appears at a particular polar angle corresponding to the Vainshtein radius.
The dip is most enhanced for the parameters near the boundary of
the region in which
the smooth matching is possible.
Clearly, this is caused by the sharp peak in $x'(r)$
at the Vainshtein radius, as seen in Fig.~\ref{fig:dxdr}.
We see from Fig.~\ref{fig:kappa.Lamb_ch} that the peak location
is certainly determined by the Vainshtein scale. From Fig.~\ref{fig:kappa.Lamb_ch} 
we also find that the depth of the dip increases as $\Lambda$ decreases.

\begin{figure}[tbp]
  \begin{center}
    \includegraphics[keepaspectratio=true,height=80mm]{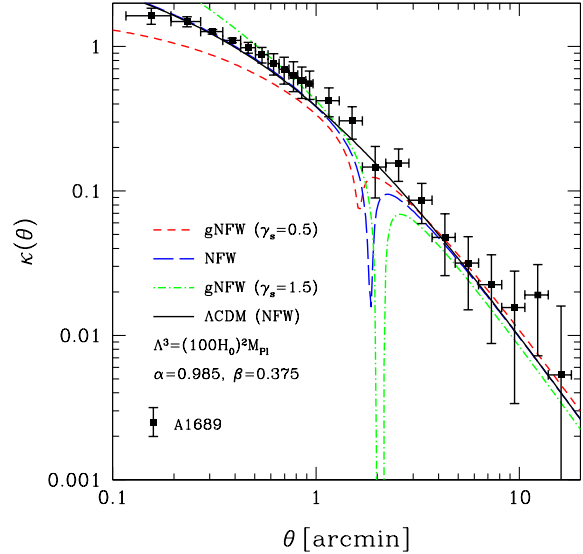}
  \end{center}
  \caption{
  The convergence $\kappa$ as a function of $\theta$ for the gNFW profile
  with $M_{\rm vir}=1.34\times 10^{15}\,\solM/h$, $c_{\rm vir}=13.8$ and $\gamma_{\rm l}=3$.
  The curves correspond to
  $\gamma_{\rm s}=0.5$ (dotted red) 
  and $\gamma_{\rm s}=1.5$ (dot-dashed green), respectively.
  For comparison, the convergence for the NFW profile ($\gamma_{\rm s}=1$)
  (with the same theory parameters) is shown by the blue dashed line.
  We take $\alpha=0.985$ and $\beta=0.375$.
}%
  \label{fig:kappagNFW}
\end{figure}

\begin{figure}[tbp]
  \begin{center}
    \includegraphics[keepaspectratio=true,height=80mm]{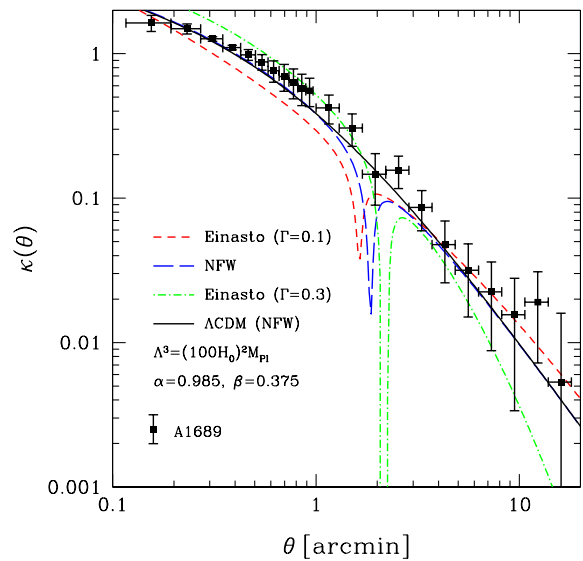}
  \end{center}
  \caption{  
  The convergence $\kappa$ as a function of $\theta$ for the Einasto profile
  with
  $M_{\rm vir}=1.34\times 10^{15}\,\solM/h$ and $r_{-2}=145\,{\rm kpc}/h$.
  The curves correspond to
  $\Gamma=0.1$ (dotted red) 
  and $\Gamma=0.3$ (dot-dashed green), respectively.
  We take $\alpha=0.985$ and $\beta=0.375$.
}%
  \label{fig:kappaEinasto}
\end{figure}

In Figs.~\ref{fig:kappagNFW} and~\ref{fig:kappaEinasto} we compare
different assumptions on the halo density profile.
Two representative profiles are considered here:
the generalized NFW (gNFW)~\cite{Zhao:1995cp, Jing:1999ir, Mandelbaum:2006pw} 
and the Einasto~\cite{Einasto, Navarro:2003ew, Gao:2007gh, Hayashi:2007uk} profiles.
We see that
a dip appears at a characteristic polar angle in the gNFW and the Einasto profiles as well.
The depth of the dip is enhanced for larger $\gamma_{\rm s}$ and larger $\Gamma$.

The appearance of a dip is expected to be a generic feature
of scalar-tensor theories exhibiting the Vainshtein mechanism,
because the essential structure of the master algebraic equation~(\ref{basic-equation})
in general cases
are the same as in massive gravity.
This helps us put constraints on scalar-tensor modification of gravity through 
the observations of cluster lensing.

The decoupling limit of massive gravity
constitutes a subclass with two free parameters in
Horndeski's theory,
which motivated us to use it for an illustrative purpose. 
We would like to point out here that there are
several caveats to be aware of when putting observational constraints
on massive gravity based on our analysis.
First, there will be some corrections to the decoupling
limit because $m$ and $1/\mpl$ are not exactly zero in reality. 
However, the corrections are small enough in the region outside the Schwarzschild radius
$r_{\rm g}$ of a lens object and inside the Compton length of the graviton~\cite{Sjors:2011iv}.
Thus, the decoupling limit can be used safely at around the Vainshtein radius,
which is relevant to our purpose, 
unless the graviton mass is so large
that the Schwarzschild radius coincides with the Compton length.
In addition, there will be other corrections since we need in fact
to consider the decoupling limit in the cosmological background\cite{deRham:2010tw}.
The corrections to
the background solution~(\ref{BG}) are expected to be
$\Delta g_{\mu\nu} = {\cal O}(H^2r^2)\eta_{\mu\nu}$ and
$\Delta \phi={\cal O}(\mpl H^2r^2)$, where $H$ is the Hubble expansion rate at the location of
the lens object.\footnote{
In the accelerating branch, the correction to
the helicity-0 mode is of order $\Lambda^3r^2$~\cite{Koyama:2011wx, deRham:2012az}. 
We do not consider this case here because the graviton mass should be small $m < H_0$
so that the expansion rate does not exceed the observed one.}  
This correction can be neglected relative to the perturbations
provided that $r < (r_{\rm g}H^{-2})^{1/3}$. This implies that
our analysis can be applied to massive gravity at
least within the Vainshtein radius $\sim (r_{\rm g}m^{-2})^{1/3}$
if $m\gtrsim H$.
Finally, and most importantly, no sensible cosmological solutions have been found so far
in massive gravity\cite{D'Amico:2011jj, DeFelice:2012mx}. 
See Refs. \cite{Gumrukcuoglu:2012aa, D'Amico:2012zv}
for recent attempts to address this issue.

\section{Conclusion}\label{sec:concl}

In this paper, we have considered static and spherically symmetric solutions 
in Horndeski's theory,
which is the most general scalar-tensor theory having second-order field equations.
Under the assumption of weak gravitational fields,
we have shown that
the problem reduces to solving an algebraic equation which is at most quintic.
Interestingly, the structure of the quintic equation
is essentially the same as the corresponding equation
derived in the context of massive gravity.
By inspecting the algebraic equation,
we have presented the conditions under which
the screened solution is obtained inside a certain radius
in the most general scalar-tensor theory.

Using our static and spherically symmetric solutions,
we have explored the possibility of 
testing modified theories of gravity endowed with the Vainshtein mechanism,
focusing on cluster lensing observations.
For an illustrative purpose and for simplicity,
we have considered a specific case of Horndeski's theory
corresponding to the decoupling limit of massive gravity,
and discussed its observational consequences.
The key effect on gravitational lensing is that the second radial
derivative of the scalar field can be
substantially large at the transition radius from screened to unscreened regions.
This results in
a dip in the convergence,
which will be a marker of the Vainshtein scale.
We have found that this dip is enhanced near the boundary
of the parameter region that allows for the screened solution.
Such a feature enables us to put constraints on modified gravity.

The following simplifications have been made in this paper:
the effect of the cosmic expansion is neglected
and the density profile is well described by the
conventional NFW, gNFW, and Einasto profiles
even in modified gravity.
However, we would like to emphasize that
the appearance of the dip in the convergence is
not dependent on the particular density profile and hence our result
is robust against the different choices of the profile.

\acknowledgments
We would like to thank
Keiichi Umetsu for providing us with his data and useful comments.
The authors thank the Yukawa Institute for Theoretical Physics at Kyoto University, 
where this work was initiated during the YITP-T-12-04 on ``Nonlinear massive gravity
theory and its observational test.''
This work was supported in part by JSPS Grant-in-Aid
for Young Scientists (B) No.~24740161 (T.K.)
and exchange visits between JSPS and DFG.
T.N. and R.S. are supported by the JSPS under Contacts
No.~23-7136 and No.~23-3430.
\appendix

\section{Stability of radial perturbations}

Let us study fluctuations propagating
around a spherically symmetric background.
For simplicity, we restrict the analysis to
radial modes and abandon the matter perturbations:
$\varphi(r)\to\varphi(r)+\delta\varphi(t, r)$,
$\Phi(r)\to\Phi(r)+\delta\Phi(t, r)$,
$\Psi(r)\to\Psi(r)+\delta\Psi(t, r)$, and $\delta T_\mu^{\;\nu}=0$.

Expanding the action to second order in perturbations
and eliminating the metric potentials by using the gravitational field equations,
we obtain the quadratic action for $\delta\varphi$:
\begin{eqnarray}
S_{\delta\varphi }=4\pi \int \D t\D r\left[
\frac{\left(r^3{\cal K}\right)'}{2}\left(\partial_ t\delta\varphi\right)^2
-r^2\frac{\partial P}{\partial x}\left(\delta\varphi'\right)^2\right].\;
\end{eqnarray}
Here we
neglected the mass term.
Instead, the above action may be deduced from
the linear equation of motion, which takes the form
$[(r^3{\cal K})'/r^2]
\partial_t^2\delta\varphi-(1/2)(\partial P/\partial x)\delta\varphi''+\cdots=0$.
The coefficient ${\cal K}$ is defined as
\begin{eqnarray}
{\cal K}(r)&=&\frac{1}{3}(\eta+6\xi^2)
+4\alpha A+2\left(\mu+6\alpha \xi+6\beta A\right)x
\nonumber\\&&
+6\left(\nu+2\alpha^2+4\beta\xi\right)x^2
+4\left(10\alpha\beta+\varpi\right)x^3
\nonumber\\&&
+30\beta^2x^4,
\end{eqnarray}
where we introduced
yet another dimensionless quantity $\varpi$
through $G_{5XX}(\phi_0, 0)=-3\varpi/\Lambda^9$.

In order to avoid the gradient instability of the radial mode,
one must require that
\begin{eqnarray}
\frac{\partial P}{\partial x}>0.
\end{eqnarray}
To avoid the ghost instability one must require that $(r^3{\cal K})'>0$.
This condition involves the new quantity $\varpi$,
which does not appear in characterizing the static and spherically symmetric solution.
Therefore, the properties of the solutions discussed in the main text
are not directly affected by this requirement.

\section{Halo profiles}\label{sec:haloprofiles}
In this appendix we summarize the halo profile models
adopted in the main text.
We briefly explain the three representative profiles:
NFW~\cite{Navarro:1995iw, Navarro:1996gj}\,,
gNFW~\cite{Zhao:1995cp, Jing:1999ir, Mandelbaum:2006pw}\,,
and Einasto~\cite{Einasto, Navarro:2003ew, Gao:2007gh, Hayashi:2007uk} profiles.
To characterize these density profiles, we introduce 
the radius $r_{-2}$\,, at which the logarithmic slope
of the density is $-2$\,, and
the virial cluster mass $M_{\rm vir}=4\pi\int^{r_{\rm vir}}_0\rho (r)r^2{\rm d}r$\,, 
which can be described by
\begin{eqnarray}
M_{\rm vir}&=&\frac{4\pi}{3}r_{\rm vir}^3 \Delta_{\rm vir} \rho_{\rm cr}(z_{\rm L}),
\end{eqnarray}
where $r_{\rm vir}$ is the virial radius, 
$\Delta_{\rm vir}$ is the virial overdensity,
and $\rho_{\rm cr}(z_{\rm L})$ is the critical density at $z_{\rm L}$\,,
where $z_{\rm L}$ denotes the redshift of the lens object.
We take $\Delta_{\rm vir}=120$ and $z_{\rm L}=0.183$ for A1689.

\subsection{NFW profile}\label{sec:NFW halo profiles}

The Navarro-Frenk-White profile is given by~\cite{Navarro:1995iw, Navarro:1996gj} 
\begin{eqnarray}
 \rho(r)=\frac{\rho_{\rm s}}{(r/r_{\rm s})(1+r/r_{\rm s})^2},
\end{eqnarray}
where  
$\rho_{\rm s}$ is the characteristic density, and $r_{\rm s}$ is 
the characteristic radius at which the slope of the density profile changes.
For the NFW profile, $r_{-2}=r_{\rm s}$.
It is useful to introduce the index of degree of concentration, 
so-called concentration parameter, $c_{\rm vir}\equiv r_{\rm vir}/r_{\rm s}$.
The virial cluster mass $M_{\rm vir}$ and the concentration parameter $c_{\rm vir}$ can 
be used as the parameters of the NFW profile 
instead of $\rho_{s}$ and $r_{s}$.

\subsection{gNFW profile}
The generalization of the NFW model may be written in
the form~\cite{Zhao:1995cp, Jing:1999ir, Mandelbaum:2006pw}
\begin{eqnarray}
\rho(r)=\frac{\rho_{\rm s}}{(r/r_{\rm s})^{\gamma_{\rm s}}(1+r/r_{\rm s})^{\gamma_{\rm l}-\gamma_{\rm s}}}. 
\label{eq:gNFW profile}
\end{eqnarray}
The NFW profile is recovered for $\gamma_{\rm s}=1$ and $\gamma_{\rm l}=3$.
We refer to the profile given by Eq.~\eqref{eq:gNFW profile} to
the generalized NFW (gNFW) profile.
For the gNFW profile, $r_{-2}=r_s(2-\gamma_{\rm s})/(\gamma_{\rm l}-2)$ and the corresponding
concentration parameter is given by
$c_{-2}\equiv r_{\rm vir}/r_{-2}=c_{\rm vir}(\gamma_{\rm l}-2)/(2-\gamma_{\rm s})$.
We can specify the gNFW profile with the virial cluster mass $M_{\rm vir}$\,,
the concentration parameter $c_{-2}$\,, and the slope indices $(\gamma_{\rm s}\,,\gamma_{\rm l})$.

\subsection{Einasto profile}
The Einasto profile is given by~\cite{Einasto, Navarro:2003ew, Gao:2007gh, Hayashi:2007uk}
\begin{eqnarray}
\rho(r)=\rho_{-2}\exp \left(-\frac{2}{\Gamma}
\left[\left(\frac{r}{r_{-2}}\right)^\Gamma-1\right]\right),
\end{eqnarray}
We can specify the Einasto profile with the virial cluster mass $M_{\rm vir}$\,,
the special radius $r_{-2}$\,, and the slope index $\Gamma$.
Based on the Millennium simulation~\cite{Springel:2005nw},
the authors of Ref.~\cite{Gao:2007gh} claimed that
cold dark matter halos can be more properly 
described by the Einasto profile than by the NFW profile.
They also argued that the best-fit
value of $\Gamma$ increases gradually with the increase of the 
virial mass, from $\Gamma\sim0.16$ for galaxy halos to 
$\Gamma\sim 0.3$ for the most massive clusters.


\end{document}